\documentclass[sigconf, preprint]{acmart}
\setcopyright{none}
\renewcommand\footnotetextcopyrightpermission[1]{}
\usepackage{multirow}
\usepackage{array}
\usepackage{xspace}
\usepackage[utf8]{inputenc} 
\usepackage[T1]{fontenc}    
\usepackage{hyperref}       
\usepackage{url}            
\usepackage{booktabs, arydshln}       
\usepackage{amsfonts}       
\usepackage{nicefrac}       
\usepackage{microtype}      
\usepackage{amsmath}
\usepackage{adjustbox}
\usepackage{bm}
\usepackage{graphicx}
\usepackage[ruled, noend]{algorithm2e}
\usepackage{float}
\usepackage{xspace}
\usepackage{siunitx}
\usepackage{dblfloatfix} 
\usepackage[caption = false]{subfig}
\usepackage{amssymb}
\usepackage{hhline}
\usepackage{caption}
\usepackage{enumitem}
\usepackage{tabularx}
\usepackage{color,colortbl}
\usepackage{hyperref}
\usepackage[toc,page]{appendix}
\usepackage[bottom]{footmisc}

\usepackage{pifont}
\newcommand{\cmark}{\ding{51}}%
\newcommand{\xmark}{\ding{55}}%

\newcommand{\method}{\textsc{Rest}\xspace}

\definecolor{gray}{HTML}{606060}
\usepackage{tikz}
\definecolor{cell}{HTML}{999999}
\definecolor{lightgray}{HTML}{F0F0F0}  
\definecolor{rowbackground}{HTML}{F9F9F9}

\newcommand{\s}[1]{\scriptsize \color{gray} $\pm$ \color{gray} #1}

\definecolor{rest-color}{HTML}{457CFC}
\definecolor{expert-color}{HTML}{319B51}
\definecolor{baseline-color}{HTML}{BDBDBD}

\newcommand{\M}[1]{\it{\textbf{#1}}}
\newcommand{\V}[1]{\it{\textbf{#1}}}

\makeatletter
\def\adl@drawiv#1#2#3{%
        \hskip.5\tabcolsep
        \xleaders#3{#2.5\@tempdimb #1{1}#2.5\@tempdimb}%
                #2\z@ plus1fil minus1fil\relax
        \hskip.5\tabcolsep}
\newcommand{\cdashlinelr}[1]{%
  \noalign{\vskip\aboverulesep
           \global\let\@dashdrawstore\adl@draw
           \global\let\adl@draw\adl@drawiv}
  \cdashline{#1}
  \noalign{\global\let\adl@draw\@dashdrawstore
           \vskip\belowrulesep}}
\makeatother

\newcommand{\hide}[1]{}

\AtBeginDocument{%
  \providecommand\BibTeX{{%
    \normalfont B\kern-0.5em{\scshape i\kern-0.25em b}\kern-0.8em\TeX}}}

\setcopyright{rightsretained}
\copyrightyear{2019}
\acmYear{2019}

\acmConference{WWW '20}{April 20 - April 24, 2020}{Taipei, Taiwan}
\acmPrice{15.00}
\acmISBN{978-1-4503-9999-9/18/06}

\begin{document}

\title{REST: Robust and Efficient Neural Networks\\ for Sleep Monitoring in the Wild}





\author{Rahul Duggal*$^1$, Scott Freitas*$^1$, Cao Xiao$^2$, Duen Horng (Polo) Chau$^1$, Jimeng Sun$^{1,3}$}

\thanks{* Both authors contributed equally to this research.}

\email{  
 {rahulduggal, safreita,polo}@gatech.edu, cao.xiao@iqvia.com, jimeng@illinois.edu
}
\affiliation{Georgia Institute of Technology$^1$, IQVIA$^2$, University of Illinois Urbana-Champaign$^3$}





\renewcommand{\shortauthors}{Duggal \& Freitas, et al.} %

\begin{abstract}

In recent years, 
significant attention has been devoted towards integrating deep learning technologies in the healthcare domain. 
However, 
to safely and practically deploy deep learning models for home health monitoring, 
two significant challenges must be addressed:
the models should be
(1) robust against noise; and 
(2) compact and energy-efficient.
We propose \method{},
a new method that simultaneously tackles both issues
via 1) \textit{adversarial training} and controlling the Lipschitz constant of the neural network through \textit{spectral regularization} while 2) enabling neural network compression through \textit{sparsity regularization}.
We demonstrate that 
\method{} produces highly-robust and efficient
models that substantially outperform the original full-sized models in the presence of noise. 
For the sleep staging task over single-channel electroencephalogram (EEG), 
the \method{} model achieves a macro-F1 score of 0.67 vs. 0.39 achieved by a state-of-the-art model in the presence of Gaussian noise while obtaining $19 \times$ parameter reduction
and $15 \times$ MFLOPS reduction on two large, real-world EEG datasets.
By deploying these models to an Android application on a smartphone,
we quantitatively observe that
\method{} allows models to achieve up to $17 \times$ energy reduction and $9 \times$ faster inference. 
We open source the code repository with this paper: \url{https://github.com/duggalrahul/REST}. 
\end{abstract}

\copyrightyear{2020}
\acmYear{2020} 
\acmConference[WWW '20]{Proceedings of The Web Conference 2020}{April 20--24, 2020}{Taipei, Taiwan} 
\acmBooktitle{Proceedings of The Web Conference 2020 (WWW '20), April 20--24, 2020, Taipei, Taiwan}
\acmPrice{}
\acmDOI{10.1145/3366423.3380241}
\acmISBN{978-1-4503-7023-3/20/04}



\maketitle

\section{Introduction}
As many as
70 million Americans suffer from sleep disorders that affects their daily functioning, long-term health and longevity. The long-term effects of sleep deprivation and sleep disorders include an increased risk of hypertension, diabetes, obesity, depression, heart attack, and stroke~\cite{altevogt2006sleep}. 
The cost of undiagnosed sleep apnea alone is estimated to exceed $100$ billion in the US \cite{american2016economic}. 

\begin{figure}[t]
    \includegraphics[width=\linewidth]{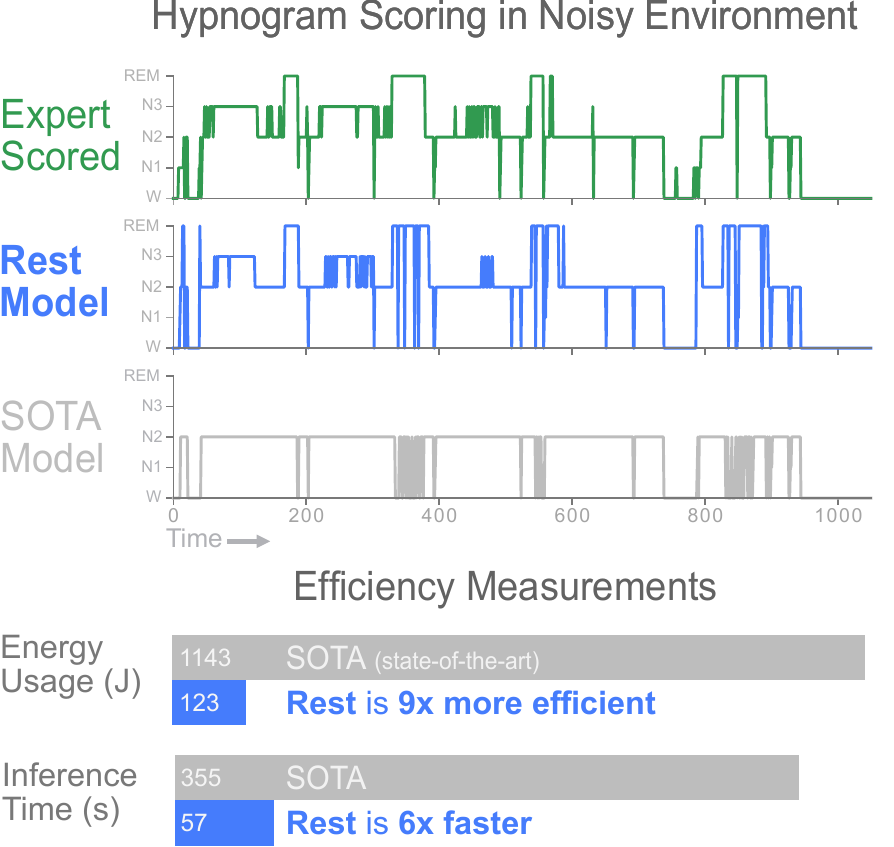}
    \caption{
    Top: we generate hypnograms for a patient in the SHHS test set. 
    In the presence of Gaussian noise, 
    our \textcolor{rest-color}{\method{}}-generated hypnogram
    closely matches the contours of the
    \textcolor{expert-color}{expert}-scored hypnogram.
    Hypnogram generated by a state-of-the-art (\textcolor{baseline-color}{SOTA}) model by Sors et al. \cite{sors2018convolutional} is considerably worse. 
    Bottom: we measure energy consumed (in Joules) and inference time (in seconds) on a smartphone to score one night of EEG recordings. \method{} is 9X more energy efficient and 6X faster than the SOTA model.
    }

    \vspace{-5mm}
    \label{fig:motivation}
\end{figure}

\begin{figure*}[t!]
    \centering
    \includegraphics[width=\textwidth]{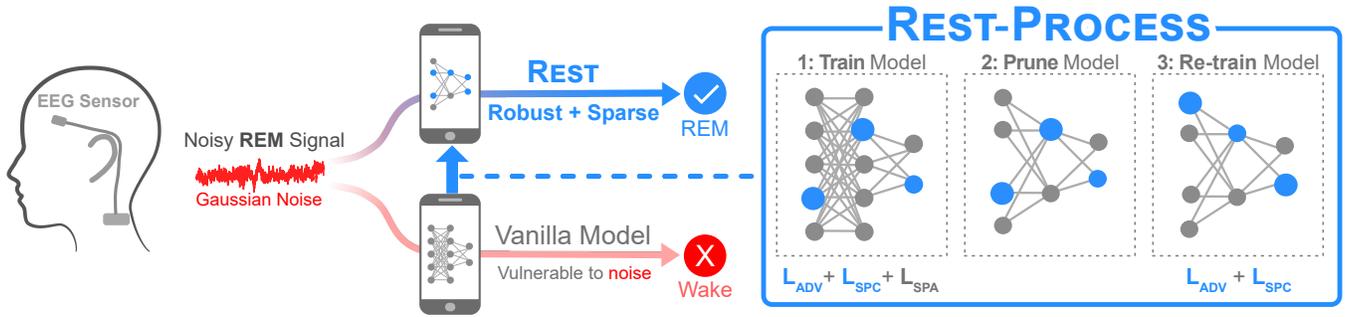}
    \caption{ \method Overview: 
    (from left) When a noisy EEG signal belonging to the REM (rapid eye movement) sleep stage enters a traditional neural network which is vulnerable to noise, it gets wrongly classified as a Wake sleep stage. On the other hand, the same signal is correctly classified as the REM sleep stage by the \method model which is both robust and sparse. (From right) \method is a three step process involving (1) training the model with adversarial training,  spectral regularization and sparsity regularization (2) pruning the model and (3) re-training the compact model.}
    \label{fig:crown}
\end{figure*}

A central tool in identifying sleep disorders is the \textbf{hypnogram}---which documents the progression of sleep stages 
(\textbf{REM} stage, \textbf{Non-REM} stages \textbf{N1} to \textbf{N3}, and \textbf{Wake} stage) over an entire night (see Fig.~\ref{fig:motivation}, top). 
The process of acquiring a hypnogram from raw sensor data is called \textbf{sleep staging}, which is the focus of this work.
Traditionally, to reliably obtain a hypnogram the patient has to undergo an overnight sleep study---called {\it polysomnography} (PSG)---at a sleep lab while wearing bio-sensors that measure physiological signals, which include electroencephalogram (EEG), 
eye movements (EOG), muscle activity or skeletal muscle activation (EMG), and heart rhythm (ECG).
The PSG data is then analyzed by a trained sleep technician and a certified sleep doctor to produce a PSG report. 
The hypnogram plays an essential role in the PSG report, where it is used to derive many important metrics such as sleep efficiency and apnea index.  
Unfortunately, manually annotating this PSG is both costly and time consuming for the doctors. 
Recent research has proposed to alleviate these issues by automatically generating the hypnogram directly from the PSG using deep neural networks \cite{biswal2017automated,supratak2017deepsleepnet}.
However, the process of obtaining a PSG report is still \textit{costly} and \textit{invasive} to  patients, reducing their participation, 
which ultimately leads to undiagnosed sleep disorders \cite{sterr2018sleep}.

One promising direction to reduce undiagnosed sleep disorders is to enable sleep monitoring at the home using commercial wearables (e.g., Fitbit, Apple Watch, Emotiv) \cite{henriksen2018using}.
However, despite significant research advances, a recent study shows that wearables using a single sensor (e.g., single lead EEG) often have lower performance for sleep staging, indicating a large room for improvement~\cite{Beattie2017-qn}.

\subsection{Contributions} 
Our contributions are two-fold---(i) we identify emerging research challenges for the task of sleep monitoring in the wild; and (ii) we propose \method{}, a novel framework that addresses these issues.

\medskip
\noindent
\textbf{I. New Research Challenges for Sleep Monitoring.}

\begin{itemize}[leftmargin=*]
\setlength\itemsep{0.5em}
  
\item \textbf{C1. Robustness to Noise.}
We observe that state-of-the-art deep neural networks (DNN) are highly susceptible to environmental noise (Fig.~\ref{fig:motivation}, top). 
In the case of wearables, noise is a serious consideration since 
bioelectrical signal sensors (e.g., electroencephalogram ``EEG'', electrocardiogram ``ECG'') are commonly susceptible to 
\textit{Gaussian} and \textit{shot} noise,
which can be introduced by electrical interferences (e.g., power-line) and user motions (e.g., muscle contraction, respiration) \cite{chang2011gaussian,blanco2008ecg,chen2010removal,bhateja2013novel}. 
This poses a need for noise-tolerant models. 
In this paper, we show that adversarial training and spectral regularization can impart significant noise robustness to sleep staging DNNs (see top of Fig~\ref{fig:motivation}).  

\item 
\textbf{C2. Energy and Computational Efficiency.}
Mobile deep learning systems have traditionally offloaded compute intensive inference to cloud servers, requiring transfer of sensitive data and assumption of available Internet.
However, this data uploading process is difficult for many healthcare scenarios because of---%
(1) {\bf privacy}: individuals are often reluctant to share health information as they consider it highly sensitive; and
(2) {\bf accessibility:} real-time home monitoring is most needed in resource-poor environments where
high-speed Internet may not be reliably available.
Directly deploying a neural network to a mobile phone bypasses these issues. However, due to the constrained computation and energy budget of mobile devices, these models need to be fast in speed and parsimonious with their energy consumption.

\end{itemize}

\noindent
\textbf{II. Noise-robust and Efficient Sleep Monitoring.} 
Having identified these two new research challenges, we propose \textbf{\method{}}, the first framework for developing noise-robust and efficient neural networks for home sleep monitoring (Fig.~\ref{fig:crown}). 
Through \method{}, our major contributions include:

\begin{itemize}[leftmargin=*]
\setlength\itemsep{0.5em}

\item 
``\textbf{R}obust and \textbf{E}fficient Neural Networks for \textbf{S}leep Moni\textbf{t}oring'' 
By integrating a novel combination of three training objectives, 
\method{} endows a model with noise robustness through 
(1) \textit{adversarial training} and 
(2) \textit{spectral regularization}; 
and promotes energy and computational efficiency by enabling compression through 
(3) \textit{sparsity regularization}.

\item \textbf{Extensive evaluation} We benchmark the performance of \method against competitive baselines, on two real-world sleep staging EEG datasets---Sleep-EDF from Physionet and Sleep Heart Health Study (SHHS). 
We demonstrate that \method{} produces highly compact models that substantially outperform the original full-sized models in the presence of noise. 
\method{} models achieves a macro-F1 score of 0.67 vs. 0.39 for the state-of-the-art model in the presence of Gaussian noise, with
$19 \times$ parameter
and $15 \times$ MFLOPS reduction. 

\item 
\textbf{Real-world deployment.}
We deploy a \method{} model onto a Pixel 2 smartphone through an Android application performing sleep staging. Our experiments reveal \method{} achieves $17 \times$ energy reduction and $9 \times$ faster inference on a smartphone, compared to uncompressed models.
\end{itemize}

\section{Related Work}
In this section we discuss related work from three areas---(1) the task of sleep stage prediction, (2) robustness of deep neural networks and (3) compression of deep learning models.

\subsection{Sleep-Stage Prediction}
Sleep staging is the task of annotating a polysomnography (PSG) report into a hypnogram, where 30 second sleep intervals are annotated into one of five sleep stages (W, N1, N2, N3, REM). 
Recently, significant effort has been devoted towards automating this annotation process using deep learning \cite{sors2018convolutional, biswal2017automated, chambon2018deep,phan2019joint, andreotti2018multichannel, dina_icml_sleep},
to name a few. While there exists a large body of research in this area---two works in particular look at both single channel \cite{biswal2017automated} and multi-channel \cite{chambon2018deep} deep learning architectures for sleep stage prediction on EEG. In \cite{biswal2017automated}, the authors develop a deep learning architecture (SLEEPNET) for sleep stage prediction that achieves expert-level accuracy on EEG data. In \cite{chambon2018deep}, the authors develop a multi-modal deep learning architecture for sleep stage prediction that achieves state-of-the-art accuracy. 
As we demonstrate later in this paper (Section~\ref{noise_robustness}), these sleep staging models are frequently susceptible to noise and suffer a large performance drop in its presence (see Figure~\ref{fig:motivation}). 
In addition, these DNNs are often overparameterized (Section~\ref{efficiency}), making deployment to mobile devices and wearables difficult. 
Through \method{}, we address these limitations and develop noise robust and efficient neural networks for edge computing.

\subsection{Noise \& Adversarial Robustness}
Adversarial robustness seeks to ensure that the output of a neural network remains unchanged under a bounded perturbation of the input; or in other words, prevent an adveresary from maliciously perturbing the data to fool a neural network.
Adversarial deep learning was popularized by \cite{goodfellow2014explaining}, where they showed it was possible to alter the class prediction of deep neural network models by carefully crafting an adversarially perturbed input. 
Since then, research suggests a strong link between adversarial robustness and noise robustness \cite{Ford2019, hendrycks2019benchmarking, tsipras2018robustness}. In particular, \cite{Ford2019} found that by performing adversarial training on a deep neural network, it becomes robust to many forms of noise (e.g., Gaussian, blur, shot, etc.). In contrast, they found that training a model on Gaussian augmented data led to models that were less robust to adversarial perturbations. We build upon this finding of adversarial robustness as a proxy for noise robustness and improve upon it through the use of spectral regularization; while simultaneously compressing the model to a fraction of its original size for mobile devices.

\subsection{Model Compression}
Model compression aims to learn a reduced representation of the weights that parameterize a neural network; shrinking the computational requirements for memory, floating point operations (FLOPS), inference time and energy.
Broadly, prior art can be classified into four directions---pruning~\cite{han2015learning}, quantization~\cite{rastegari2016xnor}, low rank approximation~\cite{xue2013restructuring} and knowledge distillation~\cite{hinton2015distilling}.
For \method, we focus on structured (channel) pruning thanks to its performance benefits (speedup, FLOP reduction) and ease of deployment with regular hardware. In structured channel pruning, the idea is to assign a measure of importance to each filter of a convolutional neural network (CNN) and achieve desired sparsity by pruning the least important ones. Prior work demonstrates several ways to estimate filter importance---magnitude of weights \cite{li2016pruning}, structured sparsity regularization \cite{wen2016learning}, regularization on activation scaling factors \cite{liu2017learning}, filter similarity \cite{cluster2019duggal} and discriminative power of filters \cite{zhuang2018discrimination}. 
Recently there has been an attempt to bridge the area of model compression with adversarial robustness through connection pruning \cite{guo2018sparse} and quantization \cite{lin2019defensive}.
Different from previous work, \method{} aims to compress a model by pruning whole filters while imparting noise tolerance through adversarial training and spectral regularization. \method{} can be further compressed through quantization \cite{lin2019defensive}.

\section{\method: Noise-Robust \& Efficient Models}\label{sec:methodology}

\method is a new method that simultaneously compresses a neural network while developing both noise and adversarial robustness.

\subsection{Overview}

Our main idea is to enable \method to endow models with these properties by 
integrating three careful modifications
of the traditional training loss function.
(1) The \textit{adversarial training} term, which builds noise robustness by training on adversarial examples (Section \ref{sec:adversarial_training}); 
(2) the \textit{spectral regularization} term, which adds to the noise robustness by constraining the Lipschitz constant of the neural network (Section \ref{sec:spectral_regularizer});
and (3) the sparsity regularization term that helps to identify important neurons and enables compression (Section \ref{sec:sparsity_regularizer}).
Throughout the paper, we follow standard notation and use capital bold letters for matrices (e.g., $\M{A}$), lower-case bold letters for vectors (e.g., $\M{a}$).

\subsection{Adversarial Training}\label{sec:adversarial_training}
The goal of adversarial training is to generate noise robustness by exposing the neural network to adversarially perturbed inputs during the training process. Given a neural network $f(\M{X};\M{W})$ with input $\M{X}$, weights $\M{W}$ and corresponding loss function $L(f(\M{X};\M{W}), y)$, adversarial training aims at solving the following min-max problem:

\begin{equation}
    \label{eq:adversarial_training}
    \underset{\M{W}}{\min} \left[ \underset{\M{X},y \sim D}{\mathbb{E}}\left( \underset{\delta \in S}{ \max}\;\; L(f(\M{X} + \delta;\M{W}),y)\right) \right]
\end{equation}

Here  $D$ is the unperturbed dataset consisting of the clean EEG signals $\M{X} \in \mathbb{R}^{K_{in} \times K_L}$ ($K_{in}$ is the number of channels and $K_L$ is the length of the signal) along with their corresponding label $y$. The inner maximization problem in (\ref{eq:adversarial_training}) embodies the goal of the adversary---that is, produce adversarially perturbed inputs (i.e., $\M{X} + \delta$) that maximize the loss function $L$. 
On the other hand, the outer minimization term aims to build robustness by countering the adversary through  minimizing the expected loss on perturbed inputs.

Maximizing the inner loss term in (\ref{eq:adversarial_training}) is equivalent to finding the adversarial signal $\M{X}_p = \M{X} + \delta$ that maximally alters the loss function $L$ within some bounded perturbation $\delta \in S$. Here $S$ is the set of allowable perturbations. Several choices exist for such an adversary. For \method, we use the iterative Projected Gradient Descent (PGD) adversary since it's one of the strongest first order attacks \cite{madry2017towards}. Its operation is described below in Equation~\ref{eq:pgd}.

\begin{equation}
\label{eq:pgd}
    \M{X}^{(t+1)}_p = \M{X}^{(t)}_p + \Pi_{\tau}\Big[\epsilon \cdot sign\Big\{\nabla_{\M{X}^{(t)}_p} L\left(f(\M{X}^{(t)}_p;\M{W}),y\right)\Big\}\Big]
\end{equation}

Here $\M{X}_p^{(0)} = \M{X}$ and at every step $t$, the previous perturbed input $\M{X}^{(t-1)}_p$ is modified with the sign of the gradient of the loss, multiplied by $\epsilon$ (controls attack strength). 
$\Pi_{\tau}$ is a function that clips the input at the positions where it exceeds the predefined $L_\infty$ bound $\tau$. 
Finally, after $n_{iter}$ iterations we have the \method{} adversarial training term $L_{adv}$ in Equation~\ref{eq:l-adv}.

\begin{equation}\label{eq:l-adv}
    L_{adv} = L(f(\M{X}^{(n_{iter})}_p;\M{W}), y)
\end{equation}

\begin{algorithm*}[!t]
\label{alg:specom}
\KwIn{Model weights $\M{W}$, EEG signal $\M{X}$ and label $y$ from dataset $\M{D}$, spectral regularization $\lambda_o$, sparsity regularization $\lambda_g$, learning rate $\alpha$, perturbation strength $\epsilon$, maximum PGD iterations $n_{iter}$ and model sparsity $s$}
\KwOut{Noise robust, compressed neural network} 
    \BlankLine
    (1) \textbf{Train the full model with \method{} loss $L_{R}$:} \\ \BlankLine
    \For{epoch = 1 to N} {
        \For{minibatch $B \subset D$} {
            \For{$\M{X} \in B$}{
                $\M{X}_p^{(1)} = \M{X}$ \\
                
                \For{k=1 to $n_{iter}$} {
                    $\M{X}^{(k+1)}_p = \M{X}_p^{(k)} + \Pi_{\tau}(\epsilon \cdot sign(\nabla_{\M{X}_p^{(k)}} L(f(\M{X}_p^{(k)};\M{W}),y)))$ 
                }
                
                $\M{W}_{grad} \leftarrow \underset{\M{X},y \sim D}{\mathbb{E}} |\triangledown_{\M{W}} L_{R}(\M{X}_{p}, y ; \M{W})|$ \\
            
                \text{where} $L_{R}= \underbrace{\vphantom{\lambda_{g} \sum_{i=1}^{N}\|\gamma^{(l)}\|_1} L(f(\M{X}_p; \M{W}), y)}_{\text{adversarial training}} + \underbrace{\lambda_{o}  \sum_{\text{layer l}=1}^{N} \|(\M{W}^{(l)})^T\M{W}^{(l)} - \M{I} \|_2}_{\text{spectral regularization}} + \underbrace{\lambda_{g} \sum_{\text{layer l}=1}^{N}\|\gamma^{(l)}\|_1}_{\text{sparsity regularization}}$
                
                $\M{W} \leftarrow \M{W} - \alpha \cdot \M{W}_{grad}$
            }
        }
    }
    \BlankLine
    (2) \textbf{Prune the trained model:}\\
    Globally prune filters from $\M{W}$ having smallest $\gamma$ values until $\frac{n_f(\M{W}')}{n_f(\M{W})} \leq s$. 
    Constrain layerwise sparsity so $\frac{n_f(\M{W}'^{(l)})}{n_f(\M{W}^{(l)})} \geq 0.1$.

    \BlankLine
    (3) \textbf{Re-train the pruned model:}\\
    Retrain compressed network $f(\M{X};\M{W}')$ using \textit{adversarial training} and \textit{spectral regularization} (no sparsity regularization).
  
 \caption{Noise Robust \& Efficient Neural Network Training (\method)}
 \label{alg:specom}
\end{algorithm*}

\subsection{Spectral Regularizer} \label{sec:spectral_regularizer}
The second term in the objective function is the spectral regularization term, which aims to constrain the change in output of a neural network for some change in input. The intuition is to suppress the amplification of noise as it passes through the successive layers of a neural network. In this section we show that an effective way to achieve this is via constraining the Lipschitz constant of each layer's weights. 

For a real valued function $f:\mathbb{R} \rightarrow \mathbb{R}$ the Lipschitz constant is a positive real value $C$ such that $|f(x_1) - f(x_2)| \leq C|x_1 - x_2|$. If $C>1$ then the change in input is magnified through the function $f$. For a neural net, this can lead to input noise amplification. On the other hand, if $C<1$ then the noise amplification effect is diminished. This can have the unintended consequence of reducing the discriminative capability of a neural net.
Therefore our goal is to set the Lipschitz constant $C=1$.
The Lipschitz constant for the $l^{th}$ fully connected layer parameterized by the weight matrix $\M{W}^{(l)} \in \mathbb{R}^{K_{in} \times K_{out}}$ is equivalent to its spectral norm $\rho(\M{W}^{(l)})$ \cite{cisse2017parseval}. Here the spectral norm of a matrix $\M{W}$ is the square root of the largest singular value of $\M{W}^T\M{W}$. The spectral norm of a 1-D convolutional layer parameterized by the tensor $\M{W}^{(l)} \in \mathbb{R}^{K_{out} \times K_{in} \times K_l}$ can be realized by reshaping it to a matrix $\M{W}^{(l)} = \mathbb{R}^{K_{out} \times (K_{in} K_l)}$ and then computing the largest singular value.

A neural network of $N$ layers can be viewed as a function $f(\cdot)$ composed of $N$ sub-functions $f(x) = f_1(\cdot) \circ f_2(\cdot) \circ ... f_N(x)$. A loose upper bound for the Lipschitz constant of $f$ is the product of Lipschitz constants of individual layers or   $\rho(f) \leq \prod_{i=1}^N \rho(f_i)$ \cite{cisse2017parseval}. The overall Lipschitz constant can grow exponentially if the spectral norm of each layer is greater than 1. On the contrary, it could go to 0 if spectral norm of each layer is between 0 and 1. Thus the ideal case arises when the spectral norm for each layer equals 1. This can be achieved in several ways \cite{yoshida2017spectral,cisse2017parseval,farnia2018generalizable}, however, one effective way is to encourage orthonormality in the columns of the weight matrix $\M{W}$ through the minimization of $\| \M{W}^T \M{W} - \M{I}\|$ where $\M{I}$ is the identity matrix. This additional loss term helps regulate the singular values and bring them close to 1. Thus we incorporate the following spectral regularization term into our loss objective, where $\lambda_{o}$ is a hyperparameter controlling the strength of the spectral regularization.

\begin{equation}
    L_{Spectral} = \lambda_{o}  \sum_{i=1}^{N} \|(\M{W}^{(i)})^T\M{W}^{(i)} - \M{I} \|_2
\end{equation}

\subsection{Sparsity Regularizer \& \method{} Loss Function}\label{sec:sparsity_regularizer}
The third term of the \method{} objective function consists of the sparsity regularizer. With this term, we aim to learn the important filters in the neural network. Once these are determined, the original neural network can be pruned to the desired level of sparsity. 

The incoming weights for filter $i$ in the $l^{th}$ fully connected (or 1-D convolutional) layer can be specified as $\V{W}^{(l)}_{i,:} \in \mathbb{R}^{K_{in}}$ (or $\V{W}^{(l)}_{i,:,:} \in \mathbb{R}^{K_{in} \times K_{L}}$). We introduce a per filter multiplicand $\gamma^{(l)}_i$ that scales the output activation of the $i^{th}$ neuron in layer $l$. By controlling the value of this multiplicand, we realize the importance of the neuron. In particular, zeroing it amounts to dropping the entire filter. Note that the $L_0$ norm on the multiplicand vector $\|\bm{\gamma}^{(l)}\|_0$, where $\bm{\gamma}^{(l)} \in \mathbb{R}^{K_{out}}$, can naturally satisfy the sparsity objective since it counts the number of non zero entries in a vector. However since the $L_0$ norm is a nondifferentiable function, we use the $L_1$ norm as a surrogate \cite{lebedev2016fast,wen2016learning,liu2017learning} which is amenable to backpropagation through its subgradient. 

To realize the per filter multiplicand $\bm{\gamma}^{(l)}_i$, we leverage the per filter multiplier within the batch normalization layer \cite{liu2017learning}. In most modern networks, a batchnorm layer immediately follows the convolutional/linear layers and implements the following operation. 

\begin{equation}
\label{eq:batchnorm}
    \bm{B}_i^{(l)} = \left(\frac{\M{A}^{(l)} - \bm{\mu}_i^{(l)}}{\bm{\sigma}_i^{(l)}}\right) \bm{\gamma}_i^{(l)} + \bm{\beta}_i^{(l)}
\end{equation}

Here $\M{A}_i^{(l)}$ denotes output activation of filter $i$ in layer $l$ while $\M{B}_i^{(l)}$ denotes its transformation through batchnorm layer $l$; $\bm{\mu}^{(l)} \in R^{K_{out}}$, $\bm{\sigma}^{(l)} \in R^{K_{out}}$ denote the mini-batch mean and standard deviation for layer $l$'s activations; and $\bm{\gamma}^{(l)} \in R^{K_{out}}$ and $\bm{\beta}^{(l)} \in R^{K_{out}}$ are learnable parameters. Our sparsity regularization is defined on $\bm{\gamma}^{(l)}$ as below, where $\lambda_{g}$ is a hyperparameter controlling the strength of sparsity regularization.

\begin{equation}
\label{eq:sparsity}
    L_{Sparsity} = \lambda_{g} \sum_{i=1}^{N}\|\bm{\gamma}^{(l)}\|_1
\end{equation}

The sparsity regularization term (\ref{eq:sparsity}) promotes learning a subset of important filters while training the model. Compression then amounts to globally pruning filters with the smallest value of multiplicands in (\ref{eq:batchnorm}) to achieve the desired model compression. Pruning typically causes a large drop in accuracy. Once the pruned model is identified, we fine-tune it via retraining.

Now that we have discussed each component of \method{}, we present the full loss function in (\ref{eq:specom}) and the training process in Algorithm~\ref{alg:specom}. A pictorial overview of the process can be seen in Figure~\ref{fig:crown}.

\begin{equation}
\begin{aligned}
\label{eq:specom}
    L_{R}= \underbrace{\vphantom{\lambda_{g} \sum_{i=1}^{N}\|\gamma^{(l)}\|_1} L(f(\M{X}_p; \M{W}), y)}_{\text{adversarial training}} &+ \underbrace{\lambda_{o}  \sum_{i=1}^{N} \|(\M{W}^{(i)})^T\M{W}^{(i)} - \M{I} \|_2}_{\text{spectral regularization}} \\ 
    &+ \underbrace{\lambda_{g} \sum_{i=1}^{N}\|\bm{\gamma}^{(l)}\|_1}_{\text{sparsity regularization}}
\end{aligned}
\end{equation}

\section{Experiments}\label{experiments}
We compare the efficacy of \method{} neural networks to four baseline models (Section~\ref{architecture})
on two publicly available EEG datasets---Sleep-EDF from Physionet \cite{goldberger2000physiobank} and Sleep Heart Health Study (SHHS) \cite{quan1997sleep}. 
Our evaluation focuses on two broad directions---\textbf{noise robustness} and \textbf{model efficiency}. Noise robustness compares the efficacy of each model when EEG data is corrupted with three types of noise: \textit{adversarial}, \textit{Gaussian} and \textit{shot}. Model efficiency compares both static (e.g., model size, floating point operations) and dynamic measurements (e.g., inference time, energy consumption). For dynamic measurements which depend on device hardware, we deploy each model to a Pixel 2 smartphone.

\begin{table}[b]
    \small
    \centering
     \begin{tabular}{lrrrrrr} 
     \toprule
     \textbf{Dataset} & \textbf{W} & \textbf{N1}  & \textbf{N2}  & \textbf{N3}(N4) & \textbf{REM} & \textbf{Total}  \\ 
     \midrule
     Sleep-EDF & 8,168 & 2,804 & 17,799 & 5,703 & 7,717 & 42,191  \\
     SHHS & 28,854 & 3,377 & 41,246 & 13,409 & 13,179 & 100,065  \\
    \bottomrule
    \end{tabular}
    \caption{Dataset summary outlining the number of 30 second EEG recordings belonging to each sleep stage class.}
    \label{tab:dataset-summary}
\end{table}

\subsection{Datasets}
Our evaluation uses two real-world sleep staging EEG datasets.

\begin{itemize}[leftmargin=*,topsep=4pt,itemsep=0ex,partopsep=0ex,parsep=1ex]
    \item \textbf{Sleep-EDF}: This dataset consists of data from two studies---age effect in healthy subjects (SC) and Temazepam effects on sleep (ST). Following \cite{supratak2017deepsleepnet}, we use whole-night polysomnographic sleep recordings on 40 healthy subjects (one night per patient) from SC. It is important to note that the SC study is conducted in the subject's homes, not a sleep center and hence this dataset is inherently noisy. However, the sensing environment is still relatively controlled since sleep doctors visited the patient's home to setup the wearable EEG sensors. After obtaining the data, the
    recordings are manually classified into one of eight classes (W, N1, N2, N3, N4, REM, MOVEMENT, UNKNOWN); we follow the steps in \cite{supratak2017deepsleepnet} and merge stages N3 and N4 into a single N3 stage and exclude MOVEMENT and UNKNOWN stages to match the five stages of sleep according to the American Academy of Sleep Medicine (AASM) \cite{berry2012aasm}.
    Each single channel EEG recording of 30 seconds corresponds to a vector of dimension $1 \times 3000$. Similar to \cite{sors2018convolutional}, while scoring at time $i$, we include EEG recordings from times $i-3, i-2, i-1, i$. Thus we expand the EEG vector by concatenating the previous three time steps to create a vector of size $1 \times 12000$.
    After pre-processing the data,
    our dataset consists of \SI{42191}{} EEG recordings, each described by a \SI{12000}{} length vector and assigned a sleep stage label from Wake, N1, N2, N3 and REM using the Fpz-Cz EEG sensor (see Table~\ref{tab:dataset-summary} for sleep stage breakdown).  
    Following standard practice \cite{supratak2017deepsleepnet}, 
    we divide the dataset on a \textit{per-patient, whole-night} basis, using 
    \SI{80}{\%} for training, 
    \SI{10}{\%} for validation,
    and
    \SI{10}{\%} for testing.
    That is, a single patient is recorded for one night and can only be in one of the three sets (training, validation, testing).
    The final number of EEG recordings in their respective splits are \SI{34820}{}, \SI{5345}{} and \SI{3908}{}. 
    While the number of recordings appear to differ from the $80$-$10$-$10$ ratio, this is because the data is split over the total number of \textit{patients}, where each patient is monitored for a time period of variable length (9 hours $\pm$ few minutes.)
    
    \item \textbf{Sleep Heart Health Study (SHHS)}:
    The Sleep Heart Health Study 
    consists of two rounds of polysomnographic recordings (SHHS-1 and SHHS-2) sampled at 125 Hz in a sleep center environment. Following \cite{sors2018convolutional}, we use only the first round (SHHS-1) containing 5,793 polysomnographic records over two channels (C4-A1 and C3-A2). Recordings are manually classified into one of six classes (W, N1, N2, N3, N4 and REM).
    As suggested in \cite{berry2012aasm}, we merge N3 and N4 stages into a single N3 stage  (see Table~\ref{tab:dataset-summary} for sleep stage breakdown). 
    We use 100 distinct patients randomly sampled from the original dataset (one night per patient). Similar to \cite{sors2018convolutional}, we look at three previous time steps in order to score the EEG recording at the current time step. This amounts to concatenating the current EEG recording of size $1 \times 3750$ (equal to 125 Hz $\times$ 30 Hz) to generate an EEG recording of size $1 \times 15000$.
    After this pre-processing, our dataset consists of \SI{100065}{} EEG recordings, each described by a \SI{15000}{} length vector and assigned a sleep stage label from the same 5 classes using the Fpz-Cz EEG sensor. 
    We use the same 80-10-10 data split as in Sleep-EDF,
    resulting in 
    \SI{79940}{} EEG recordings for training, \SI{9999}{} for validation,
    and 
    \SI{10126}{} for testing.
    
\end{itemize}

\subsection{Model Architecture and Configurations}\label{architecture}
We use the sleep staging CNN architecture proposed by \cite{sors2018convolutional}, since it achieves state-of-the-art accuracy for sleep stage classification using single channel EEG.
We implement all models in PyTorch 0.4. 
For training and evaluation, we use a server equipped with an Intel Xeon E5-2690 CPU, 250GB RAM and 8 Nvidia Titan Xp GPUs. 
Mobile device measurements use a Pixel 2 smartphone with an Android application running Tensorflow Lite\footnote{TensorFlow Lite: \url{https://www.tensorflow.org/lite}}.
With \cite{sors2018convolutional} as the architecture for all baselines below, 
we compare the following 6 configurations:

\vspace{2mm}
\begin{enumerate}[leftmargin=*]
    \item \textbf{Sors} \cite{sors2018convolutional}: Baseline neural network model trained on unperturbed data. This model contains 12 1-D convolutional layers followed by 2 fully connected layers and achieves state-of-the-art performance on sleep staging using single channel EEG.
    
    \item \textbf{Liu} \cite{liu2017learning}: We train on unperturbed data and compress the Sors model using sparsity regularization as proposed in \cite{liu2017learning}. 
    
    \item \textbf{Blanco} \cite{blanco1997applying}: We use same setup from Liu above.
    During test time, the noisy test input is filtered using a bandpass filter with cutoff 0.5Hz-40Hz
    This technique is commonly used for removing noise in EEG analysis \cite{blanco1997applying}.
    
    \item \textbf{Ford} \cite{Ford2019}: We train and compress the Sors model with sparsity regularization on input data perturbed by Gaussian noise.  
    Gaussian training parameter $c_g$ = 0.2 controls the perturbation strength during training;
    identified through a line search in Section~\ref{sec:hyperparam-select}. 
    
    \item \textbf{\method(A)}: Our compressed Sors model obtained through adversarial training and sparsity regularization. We use the hyperparameters: $\epsilon$ = 10, $ n_{iter}$= 5/10 (SHHS/Sleep-EDF), where $\epsilon$ is a \textit{key} variable controlling the strength of adversarial perturbation during training. The optimal $\epsilon$ value is determined through a line search described in Section~\ref{sec:hyperparam-select}. 
    
    \item \textbf{\method(A+S)}: Our compressed Sors model obtained through adversarial training, spectral and sparsity regularization. We set the spectral regularization parameter $\lambda_o$ = $3 \times 10^{-3}$ and sparsity regularization parameter $\lambda_g$ = $10^{-5}$ based on a grid search in Section~\ref{sec:hyperparam-select}. 
    
\end{enumerate}
\vspace{2mm}

All models are trained for 30 epochs using SGD. 
The initial learning rate is set to 0.1 and multiplied by 0.1 at epochs 10 and 20;
the weight decay is set to 0.0002. 
All compressed models use the same compression method, consisting of weight pruning followed by model re-training. The sparsity regularization parameter $\lambda_g = 10^{-5}$ is identified through a grid search with $\lambda_o$ (after determining $\epsilon$ through a line search). Detailed analysis of the hyperparameter selection for $\epsilon$, $\lambda_o$ and $\lambda_g$ can be found in Section~\ref{sec:hyperparam-select}. Finally, we set a high sparsity level $s$ = 0.8 (80\% neurons from the original networks were pruned) after observation that the models are overparametrized for the task of sleep stage classification.

\subsection{Evaluation Metrics}
\medskip
\noindent
\textbf{Noise robustness metrics} 
To study the noise robustness of each model configuration, 
we evaluate macro-F1 score in the presence of
three types of noise: 
adversarial, Gaussian and shot. 
We select macro-F1 since it is a standard metric for evaluating classification performance in imbalanced datasets.
Adversarial noise is defined at three strength levels through $\epsilon = 2/6/12$ in Equation~\ref{eq:pgd}; Gaussian noise at three levels through $c_g = 0.1/0.2/0.3$ in Equation~\ref{eq:gaussian}; and shot noise at three levels through $c_s = 5000/2500/1000$ in Equation~\ref{eq:shot}. These parameter values are chosen based on prior work \cite{madry2017towards,hendrycks2019benchmarking} and empirical observation. 
For evaluating robustness to adversarial noise, we assume the white box setting where the attacker has access to model weights.
The formulation for Gaussian and shot noise is in Equation~\ref{eq:gaussian} and \ref{eq:shot}, respectively.
    
\begin{equation}\label{eq:gaussian}
    \M{X}_{gauss} = \M{X} + N(0, c_g \cdot \sigma_{train})
\end{equation}

In Equation~\ref{eq:gaussian}, $\sigma_{train}$ is the standard deviation of the training data and $N$ is the normal distribution. The noise strength---low, medium and high---corresponds to $c_g = 0.1/0.2/0.3$.

\begin{equation}\label{eq:shot}
\begin{aligned}
    &\M{X}_{norm} = \frac{\M{X} - x_{min}}{x_{max} - x_{min} } \\
    &\M{X}' = clip_{0,1}\left(\frac{Poisson(\M{X}_{norm}.c_s)}{c_s}\right) \\
    &\M{X}_{shot} = \M{X}' . (x_{max} - x_{min}) + x_{min} \\
\end{aligned}
\end{equation}

In Equation~\ref{eq:shot}, $x_{min},x_{max}$ denote the minimum and maximum values in the training data; and $clip_{0,1}$ is a function that projects the input to the range [0,1].

\medskip
\noindent
\textbf{Model efficiency metrics} 
To evaluate the efficiency of each model configuration, we use the following measures:

\begin{itemize}[leftmargin=*,topsep=0pt,itemsep=0ex,partopsep=2ex,parsep=1ex]

    \item \textbf{Parameter Reduction}: Memory consumed (in KB) for storing the weights of a model.
    
     \item \textbf{Floating point operations (FLOPS)}: Number of multiply and add operations performed by the model in one forward pass. Measurement units are Mega ($10^6$).
    
    \item \textbf{Inference Time}: Average time taken (in seconds) to score one night of EEG data. We assume a night consists of 9 hours and amounts to 1,080 EEG recordings (each of 30 seconds). This is measured on a Pixel 2 smartphone.
    
    \item \textbf{Energy Consumption}: Average energy consumed by a model (in Joules) to score one night of EEG data on a Pixel 2 smartphone. To measure consumed energy, we implement an infinite inference loop over EEG recordings 
    until the battery level drops from $100\%$ down to $85\%$.
    For each unit percent drop (i.e., 15 levels), we log the number of iterations $N_i$ performed by the model. Given that a standard Pixel 2 battery can deliver 2700 mAh at 3.85 Volts, we use the following conversion to estimate energy consumed $E$ (in Joules) for a unit percent drop in battery level $E = \frac{2700}{1000} \times 3600 \times 3.85$. The total energy for inferencing over an entire night of EEG recordings is then calculated as $\frac{E}{N_{i}} \times 1080$ where $N_i$ is the number of inferences made in the unit battery drop interval. We average this for every unit battery percentage drop from $100\%$ to $85\%$ (i.e., 15 intervals) to calculate the average energy consumption
\end{itemize}

\subsection{Hyperparameter Selection}\label{sec:hyperparam-select}

Optimal hyper-parameter selection is crucial for obtaining good performance with both baseline and \method{} models. We systematically conduct a series of line and grid searches to determine ideal values of $\epsilon$, $c_g$, $\lambda_o$ and $\lambda_g$ using the validation sets. 

\begin{figure}[!t]
    \centering
    \includegraphics[width=0.9\linewidth]{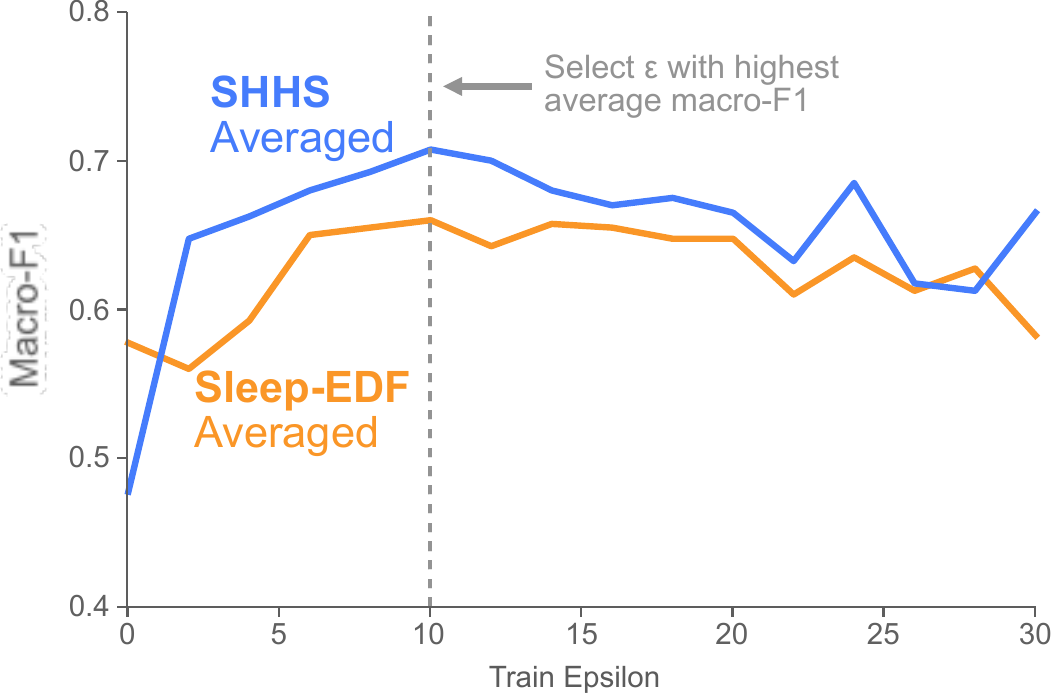}
    \caption{Line search results for $\epsilon$ on Sleep-EDF and SHHS datasets. 
    We select $\epsilon$=10, since it provides the best average macro-F1 score on both datasets.
    }
    \label{fig:train-epsilon}
\end{figure}

\medskip
\noindent
\textbf{Selecting $\epsilon$\hspace{0.2cm}}
This parameter controls the perturbation strength of adversarial training in Equation~\ref{eq:pgd}. Correctly setting this parameter is critical since a small $\epsilon$ value will have no effect on noise robustness, while too high a value will lead to poor benign accuracy.  
We follow standard procedure and determine the optimal $\epsilon$ on a per-dataset basis \cite{madry2017towards}, conducting a line search across $\epsilon \in [$0,30$]$ in steps of 2.
For each value of $\epsilon$ we measure benign and adversarial validation macro-F1 score, where adversarial macro-F1 is an average of three strength levels: low ($\epsilon$=2), medium ($\epsilon$=6) and high ($\epsilon$=12). 
We then select the $\epsilon$ with highest macro-F1 score averaged across the benign and adversarial macro-F1. 
Line search results are shown in Figure~\ref{fig:train-epsilon}; we select $\epsilon=10$ for both dataset since it's the value with highest average macro-F1.

\begin{table}[!t]
    \setlength{\tabcolsep}{5.7pt}
    \centering
     \begin{tabular}{crrrrrr} 
     \toprule
     & & & \multicolumn{3}{c}{{\textbf{Guassian F1}}} & \\
    \cmidrule(l){4-6} 
     
    & \textbf{$c_g$} &  \textbf{Benign F1} & \textbf{Low} & \textbf{Med} & \textbf{High} & \textbf{Average F1} \\
    \midrule
    
    \multirow{3}{*}{\rotatebox[origin=c]{90}{\textbf{EDF}}}
    & 0.1 & 0.75 & 0.76 & 0.7 & 0.5 & 0.68 \\
    &  0.2 & 0.7 & 0.72 & 0.75 & 0.64 & \textbf{0.70} \\
    & 0.3 & 0.67 & 0.68 & 0.71 & 0.75 & 0.7025 \\
     
    \midrule

    \multirow{3}{*}{\rotatebox[origin=c]{90}{\textbf{SHHS}}}
    & 0.1 & 0.69 & 0.74 & 0.45 & 0.21 & 0.52 \\
    & 0.2 & 0.68 & 0.69 & 0.68 & 0.43 & \textbf{0.62} \\
    & 0.3 & 0.55 & 0.57 & 0.65 & 0.74 & 0.63 \\
    \bottomrule
    \end{tabular}
    \caption{Line search results for identifying optimal $c_g$ on Sleep-EDF and SHHS datasets. Macro-F1 is abbreviated F1 in table; average macro-F1 is the mean of all macro-F1 scores.
    We select $c_g$=0.2 for both datasets as it represents a good trade-off between benign and Gaussian macro-F1.
    }
    \label{tab:line-search-gaussian}
\end{table}

\begin{table}[!h]
\setlength{\tabcolsep}{5.3pt}
    \centering
     \begin{tabular}{rrrrrrr} 
     \toprule
     & & & \multicolumn{3}{c}{{\textbf{Adversarial F1}}} & \\
    \cmidrule(l){4-6} 
     
     \textbf{$\lambda_o$} & \textbf{$\lambda_g$} & \textbf{Benign F1} & \textbf{Low} & \textbf{Med} & \textbf{High} & \textbf{Avg. F1} \\
     \midrule
     
    0.001 & 1E-04 & 0.73 & 0.66 & 0.65 & 0.61 & 0.66 \\
    0.003 & 1E-04 & 0.72 & 0.64 & 0.63 & 0.59 & 0.65 \\
    0.005 &	1E-04 & 0.72 & 0.65 & 0.64 & 0.62 & 0.66 \\
    0.001 &	1E-05 &	0.73 & 0.66 & 0.65 & 0.62 & 0.67 \\
    \rowcolor{lightgray} 
    0.003 & 1E-05 & 0.73 & 0.67 & 0.66 & 0.62 & \textbf{0.67} \\
    0.005 &	1E-05 & 0.73 & 0.64 & 0.64 & 0.62 & 0.66 \\
    \bottomrule
    \end{tabular}
    \captionof{table}{Grid search results for $\lambda_o$ and $\lambda_g$ on Sleep-EDF dataset. 
    Macro-F1 is abbreviated as F1 in table; average macro-F1 is the mean of all macro-F1 scores.
    We select $\lambda_o$ and $\lambda_g$ with highest average macro-F1 score.
    }
    \label{tab:grid-search-lambdas}
\end{table}

\medskip
\noindent
\textbf{Selecting $c_g$\hspace{0.2cm}} 
This parameter controls the noise perturbation strength of Gaussian training in Equation~\ref{eq:gaussian}.
Similar to $\epsilon$, we determine $c_g$ on a per-dataset basis, conducting a line search across $c_g$ values: 0.1 (low), 0.2 (medium) and 0.3 (high). 
Based on results from Table~\ref{tab:line-search-gaussian}, we select $c_g$=0.2 for both datasets since it provides the best average macro-F1 score while minimizing the drop in benign accuracy. 

\medskip
\noindent
\textbf{Selecting $\lambda_o$ and $\lambda_g$ \hspace{0.2cm}} 
These parameters determine the strength of spectral and sparsity regularization in Equation~\ref{eq:specom}.
We determine the best value for $\lambda_o$ and $\lambda_g$ through a grid search across the following parameter values $\lambda_o=[0.001, 0.003, 0.005]$ and $\lambda_g=[1E-04, 1E-05]$. 
Based on results from Table~\ref{tab:grid-search-lambdas}, we select $\lambda_o=0.003$ and $\lambda_g=1E-05$.
Since these are model dependent parameters, we calculate them once on the Sleep-EDF dataset and re-use them for SHHS.

\begin{table*}[t!]
\setlength{\tabcolsep}{2.5pt}
\centering

 \begin{tabular}{clcrrrrrrrrrr }
 \toprule
 & & & & \multicolumn{3}{c}{\textbf{Adversarial}} & \multicolumn{3}{c}{\textbf{Gaussian}} & \multicolumn{3}{c}{\textbf{Shot}} \\
\cmidrule(l){5-7} \cmidrule(l){8-10} \cmidrule(l){11-13} 
 
 \textbf{Data} & \textbf{Method}  & \textbf{Compress} & \textbf{No noise} & Low & Med & High & Low & Med & High & Low & Med & High \\ 
 \midrule
 \multirow{6}{*}{\rotatebox[origin=c]{90}{\textbf{Sleep-EDF}}}
 
 \vspace{2mm} & Sors \cite{sors2018convolutional} & \xmark & 0.67 \s{0.02} & 0.57 \s{0.02} & 0.51 \s{0.04} & 0.19 \s{0.06} & 0.66 \s{0.03} & 0.60 \s{0.03} & 0.39 \s{0.08} & 0.58 \s{0.04} & 0.42 \s{0.08} & 0.11 \s{0.03} \\ 
 
 & Liu \cite{liu2017learning} & \cmark & \textbf{0.69} \s{0.02} & 0.52 \s{0.07} & 0.41 \s{0.07} & 0.09 \s{0.02} & 0.67 \s{0.02} & 0.53 \s{0.02} & 0.28 \s{0.04} & 0.52 \s{0.03} & 0.31 \s{0.04} & 0.06 \s{0.01} \\ 
 & Blanco \cite{blanco1997applying} & \cmark & 0.68 \s{0.01} & 0.51 \s{0.06} & 0.40 \s{0.06} & 0.09 \s{0.02} & 0.65 \s{0.02} & 0.54 \s{0.04} & 0.31 \s{0.10} & 0.53 \s{0.04} & 0.34 \s{0.09} & 0.08 \s{0.02} \\ 
 \vspace{2mm} & Ford \cite{Ford2019} & \cmark & 0.64 \s{0.01} & 0.59 \s{0.01} & 0.60 \s{0.02} & 0.31 \s{0.08} & 0.65 \s{0.01} & 0.67 \s{0.02} & 0.57 \s{0.03} & 0.67 \s{0.02} & 0.60 \s{0.02} & 0.10 \s{0.01} \\ 

 & \method(A) & \cmark & 0.66 \s{0.02} & 0.64 \s{0.02} & 0.64 \s{0.02} & 0.61 \s{0.02} & 0.66 \s{0.02} & 0.67 \s{0.01} & 0.66 \s{0.01} & 0.67 \s{0.01} & 0.66 \s{0.01} & \textbf{0.42} \s{0.06} \\ 
 & \method(A+S) & \cmark & \textbf{0.69} \s{0.01} & \textbf{0.67} \s{0.02} & \textbf{0.66} \s{0.01} & \textbf{0.61} \s{0.03} & \textbf{0.69} \s{0.01} & \textbf{0.68} \s{0.01} & \textbf{0.67} \s{0.02} & \textbf{0.68} \s{0.01} & \textbf{0.67} \s{0.02} & \textbf{0.42} \s{0.08} \\ 
 
 \midrule
 
 \multirow{6}{*}{\rotatebox[origin=c]{90}{\textbf{SHHS}}}
 \vspace{2mm} & Sors \cite{sors2018convolutional} & \xmark & \textbf{0.78} \s{0.01} & 0.62 \s{0.03} & 0.46 \s{0.03} & 0.33 \s{0.00} & 0.64 \s{0.03} & 0.43 \s{0.02} & 0.35 \s{0.04} & 0.69 \s{0.02} & 0.59 \s{0.03} & 0.45 \s{0.01} \\ 
 
 & Liu \cite{liu2017learning} & \cmark & 0.77 \s{0.01} & 0.61 \s{0.02} & 0.49 \s{0.04} & 0.34 \s{0.03} & 0.66 \s{0.05} & 0.45 \s{0.05} & 0.34 \s{0.04} & 0.70 \s{0.04} & 0.62 \s{0.04} & 0.47 \s{0.05} \\ 
 & Blanco \cite{blanco1997applying} & \cmark & 0.77 \s{0.01} & 0.60 \s{0.03} & 0.47 \s{0.04} & 0.33 \s{0.02} & 0.64 \s{0.07} & 0.43 \s{0.05} & 0.34 \s{0.04} & 0.67 \s{0.06} & 0.59 \s{0.05} & 0.46 \s{0.04} \\ 
 \vspace{2mm}& Ford \cite{Ford2019} & \cmark & 0.62 \s{0.02} & 0.59 \s{0.01} & 0.62 \s{0.00} & 0.59 \s{0.05} & 0.66 \s{0.00} & 0.75 \s{0.04} & 0.47 \s{0.10} & 0.65 \s{0.00} & 0.68 \s{0.01} & 0.74 \s{0.04} \\ 

 & \method(A) & \cmark & 0.70 \s{0.01} & 0.68 \s{0.00} & 0.70 \s{0.01} & 0.67 \s{0.01} & 0.72 \s{0.01} & 0.76 \s{0.01} & 0.58 \s{0.03} & 0.72 \s{0.01} & 0.74 \s{0.01} & 0.76 \s{0.01}  \\ 
 & \method(A+S) & \cmark & 0.72 \s{0.01} & \textbf{0.69} \s{0.01} & \textbf{0.70} \s{0.01} & \textbf{0.69} \s{0.02} & \textbf{0.74} \s{0.01} & \textbf{0.77} \s{0.01} & \textbf{0.62} \s{0.03} & \textbf{0.73} \s{0.01} & \textbf{0.75} \s{0.01} & \textbf{0.78} \s{0.00} \\ 
 
\bottomrule
\end{tabular}
\caption{
Meta Analysis: Comparison of macro-F1 scores achieved by each model. The models are evaluated on Sleep-EDF and SHHS datasets with three types and strengths of noise corruption. 
We bold the compressed model with the best performance (averaged over 3 runs) and report the standard deviation of each model next to the macro-F1 score. 
\method{} performs better in \textit{all} noise test measurements. 
}
\label{tab:noise-robustness}
\end{table*}

\subsection{Noise Robustness}\label{noise_robustness}
To evaluate noise robustness,  we ask the following questions---(1) what is the impact of \method{} on model accuracy with and without noise in the data? and (2) how does \method{} training compare to baseline methods of benign training, Gaussian training and noise filtering?
In answering these questions, we analyze noise robustness of models at three scales:
(i) meta-level macro-F1 scores; 
(ii) meso-level confusion matrix heatmaps; 
and 
(iii) granular-level single-patient hypnograms.

\medskip
\noindent
\textbf{I. Meta analysis: Macro-F1 Scores} \label{subsec:aggregated_results}
In Table~\ref{tab:noise-robustness}, we present a high-level overview of model performance through macro-F1 scores on three types and strength levels of noise corruption.
The Macro-F1 scores and standard deviation are reported by averaging over three runs for each model and noise level. We identify multiple key insights as described below:

\begin{enumerate}[topsep=2mm, itemsep=0mm, parsep=1mm, leftmargin=*]

    \item \textbf{\method Outperforms Across All Types of Noise}  As demonstrated by the higher macro-F1 scores, \method{} outperforms all baseline methods in the presence of noise. In addition, \method{} has a low standard deviation, indicating model performance is not dependent on weight initialization.
    
    \item \textbf{Spectral Regularization Improves Performance}  \method$(A+S)$ consistently improves upon \method$(A)$, indicating the usefulness of spectral regularization towards enhancing noise robustness by constraining the Lipschitz constant.
    
    \item \textbf{SHHS Performance Better Than Sleep-EDF} Performance is generally better on the SHHS dataset compared to Sleep-EDF. One possible explanation is due to the SHHS dataset being less noisy in comparison to the Sleep-EDF dataset. This stems from the fact that the SHHS study was performed in the hospital setting while Sleep-EDF was undertaken in the home setting.
    
    \item \textbf{Benign \& Adversarial Accuracy Trade-off} Contrary to the traditional trade-off between benign and adversarial accuracy, \method{} performance matches Liu in the no noise setting on sleep-EDF. 
    This is likely attributable to the noise in the Sleep-EDF dataset, which was collected in the home setting. 
    On the SHHS dataset, the Liu model outperforms \method{} in the no noise setting, where data is captured in the less noise prone hospital setting.
    Due to this, \method{} models are best positioned for use in noisy environments (e.g., at home); while traditional models are more effective in controlled environments (e.g., sleep labs).
\end{enumerate}

\begin{figure*}[t!]
    \centering
    \includegraphics[width=\textwidth]{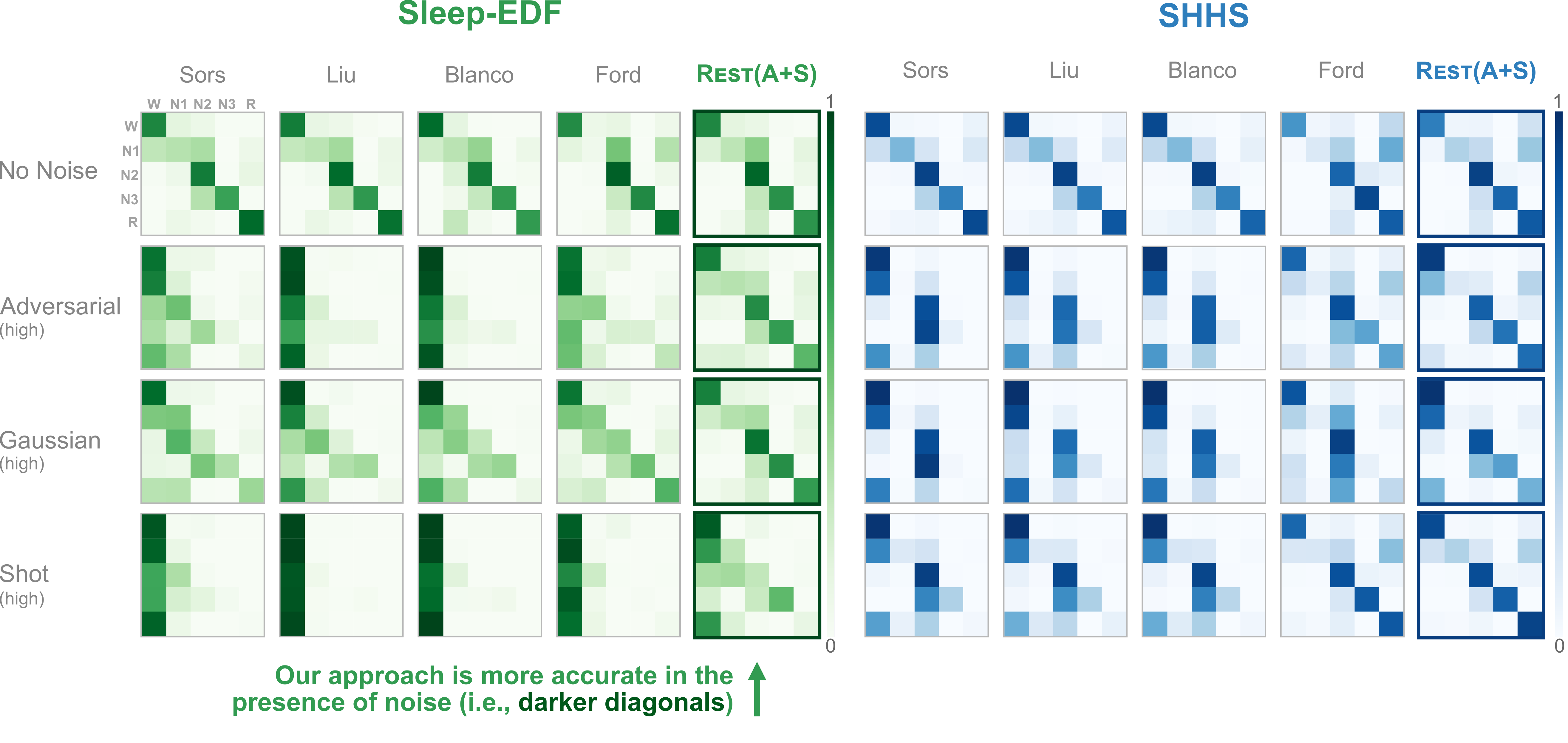}
    \captionof{figure}{
    Meso Analysis: Class-wise comparison of model predictions.
    The models are evaluated over the SHHS test set perturbed with different noise types. 
    In each confusion matrix, 
    rows are ground-truth classes
    while columns are predicted classes. 
    The intensity of a cell is obtained by normalizing the score with respect to the class membership. 
    When a cell has a value of 1 (dark blue, or dark green) the model predicts every example correctly, the opposite occurs at 0 (white). 
    A model that is performing well would have a dark diagonal and light off-diagonal. 
    \method{} has the darkest cells along the diagonal on both datasets.}
    \label{fig:heatmap}
    \vspace{0.25cm}
\end{figure*}

\medskip
\noindent
\textbf{II. Meso Analysis: Per-class Performance} \label{subsec:confusion_matrix}
We visualize and identify class-wise trends using confusion matrix heatmaps (Fig.~\ref{fig:heatmap}).
Each confusion matrix describes a model's performance for a given level of noise (or no noise).
A model that is performing well should have a dark diagonal and light off-diagonal.
We normalize the rows of each confusion matrix to accurately represent class predictions in an imbalanced dataset. 
When a matrix diagonal has a value of 1 (dark blue, or dark green) the model predicts every example correctly;
the opposite occurs at 0 (white). 
Analyzing Figure~\ref{fig:heatmap}, we identify the following key insights:

\begin{enumerate}[topsep=2mm, itemsep=0mm, parsep=1mm, leftmargin=*]
    \item \textbf{\method{} Performs Well Across All Classes}  \method{} accurately predicts each sleep stage (W, N1, N2, N3, REM) across multiple types of noise (Fig.~\ref{fig:heatmap}, bottom 3 rows), as evidenced by the dark diagonal.
    In comparison, each baseline method has considerable performance degradation (light diagonal) in the presence of noise. 
    This is particularly evident on the Sleep-EDF dataset (left half) where data is collected in the noisier home environment.
    
    \item \textbf{N1 Class Difficult to Predict} 
    When no noise is present (Fig.~\ref{fig:heatmap}, top row), each method performs well as evidenced by the dark diagonal, except on the N1 sleep stage class.
    This performance drop is likely due to the limited number of N1 examples in the datasets (see Table~\ref{tab:dataset-summary}).

    \item \textbf{Increased Misclassification Towards ``Wake'' Class} 
    On the Sleep-EDF dataset, shot and adversarial noise cause the baseline models to mispredict classes as Wake. 
    One possible explanation is that the models misinterpret the additive noise as evidence for the wake class which has characteristically large fluctuations.
\end{enumerate}

\begin{figure*}[b!]
    \centering
    \includegraphics[width=\textwidth]{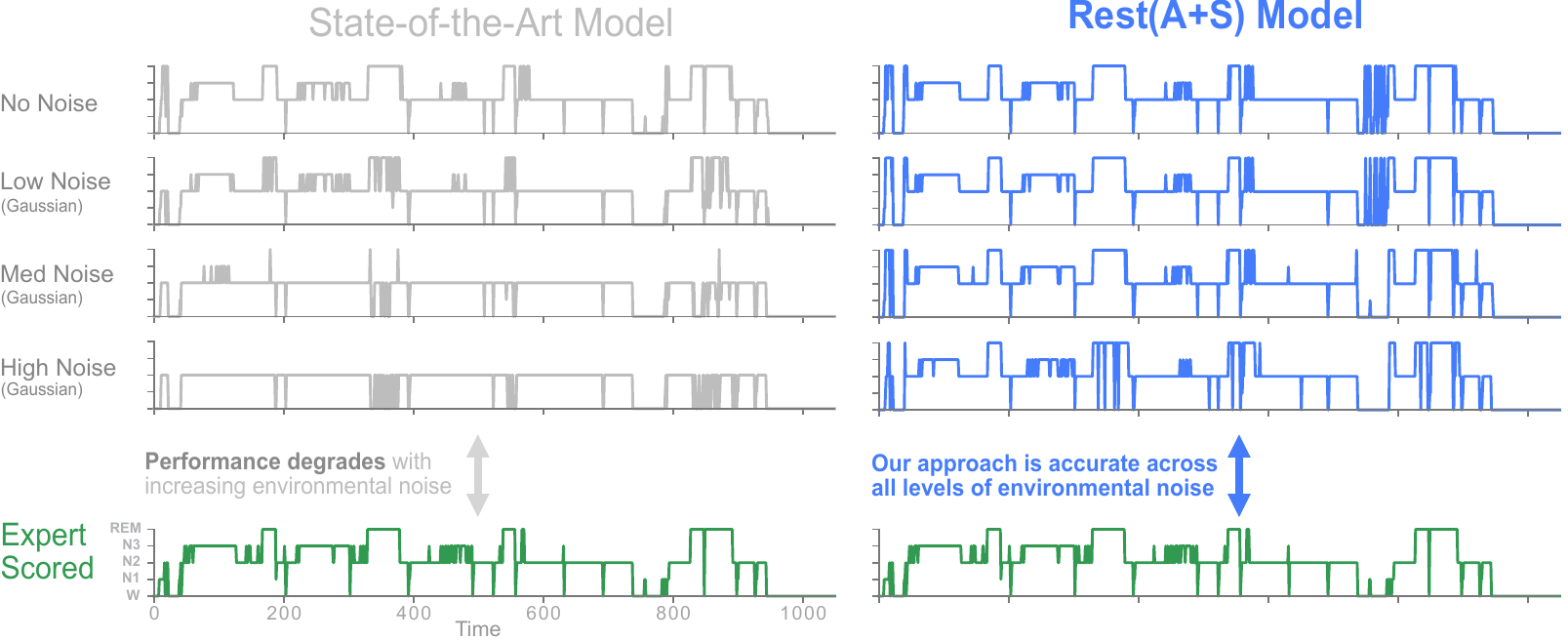}
    \captionof{figure}{Granular Analysis: Comparison 
    of the overnight hypnograms obtained for a patient in the SHHS test set. 
    The hypnograms are generated using the Sors (left) and \method (right) models in the presence of increasing strengths of Gaussian noise. 
    When no noise is present (top row), both models perform well, closely matching the ground truth (bottom row). 
    However, with increasing noise, Sors performance rapidly degrades, while \method{} continues to generate accurate hypnograms.}
    \label{fig:hypnogram-full}
\end{figure*}

\noindent
\textbf{III. Granular Analysis: Single-patient Hypnograms} \label{subsec:hypnogram}
We want to more deeply understand how our \method{} models counteract noise at the hypnogram level.
Therefore, we select a test set patient from the SHHS dataset, and generate and visualize the patient's overnight hypnograms using the Sors and \method{} models on three levels of Gaussian noise corruption (Figure~\ref{fig:hypnogram-full}).
Each of these hypnograms is compared to a trained technicians hypnogram (expert scored in Fig.~\ref{fig:hypnogram-full}), representing the ground-truth. 
We inspect a few more test set patients using the above approach, and identify multiple key representative insights:

\begin{enumerate}[topsep=2mm, itemsep=0mm, parsep=1mm, leftmargin=*]
    \item \textbf{Noisy Environments Require Robust Models} As data noise increases, Sors performance degrades. This begins at the low noise level, further accelerates in the medium level and reaches nearly zero at the high level.
    In contrast, \method{} effectively handles all levels of noise, generating an accurate hypnogram at even the highest level. 
    
    \item \textbf{Low Noise Environments Give Good Performance} In the no noise setting (top row) both the Sors and \method{} models generate accurate hypnograms, closely matching the contours of expert scoring (bottom).

\end{enumerate}

\begin{figure}[t!]
    \centering
    \includegraphics[width=0.38\textwidth]{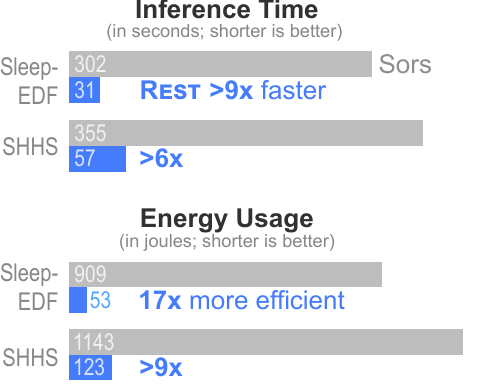}
    \captionof{figure}{Time and energy consumption for scoring a single night of EEG recordings. \method{}(A+S) is significantly faster and more energy efficient than the state-of-the-art Sors model. Evaluations were done on a Pixel 2 smartphone.}
    \label{fig:efficiency}
\end{figure}

\subsection{Model Efficiency}
\label{efficiency}
We measure model efficiency along two dimensions---%
(1) \textit{static metrics}: amount of memory required to store weights in memory and FLOPS; and 
(2) \textit{dynamic metrics}: inference time and energy consumption. 
For dynamic measurements that depend on device hardware, we deploy each model to a Pixel 2 smartphone.

\smallskip
\noindent \textbf{Analyzing Static Metrics: Memory \& Flops} 
Table \ref{tab:efficiency} describes the size (in KB) and computational requirements (in MFlops) of each model. We identify the following key insights:

\begin{enumerate}[topsep=2mm, itemsep=0mm, parsep=1mm, leftmargin=*]

    \item \textbf{\method{} Models Require Fewest FLOPS} On both datasets, \method{} requires the least number of FLOPS.
    
    \item \textbf{\method{} Models are Small} \method{} models are also smaller (or comparable) to baseline compressed models while achieving significantly better noise robustness. 
    \item \textbf{Model Efficiency and Noise Robustness} Combining the insights from Section~\ref{noise_robustness} and the above, we observe that \method{} models have significantly better noise robustness while maintaining a competitive memory footprint. This suggests that robustness is more dependent on the the training process, rather than model capacity.
    
\end{enumerate}

\begin{table}[t!]
    \setlength{\tabcolsep}{14pt}
    \centering
     \begin{tabular}{c|lrr} 
     \toprule
     \textbf{Data} & \textbf{Model} & \textbf{Size} (KB)  & \textbf{MFlops} \\ 
     \midrule
     \multirow{6}{*}{\rotatebox[origin=c]{90}{\textbf{Sleep-EDF}}}
     & Sors~\cite{sors2018convolutional} & 8,896 & 1451 \\
     & Liu~\cite{liu2017learning} & \textbf{440} & 127 \\
     & Blanco~\cite{blanco1997applying} & 440 & 127 \\
     & Ford~\cite{Ford2019} & 448 & 144 \\ 
     & \method(A) & 464 & 98 \\ 
     & \method(A+S) & 449 & \textbf{94} \\ 
     
      \midrule
     
     \multirow{6}{*}{\rotatebox[origin=c]{90}{\textbf{SHHS}}}
     & Sors~\cite{sors2018convolutional} & 8,996 & 1815 \\
     & Liu~\cite{liu2017learning} & \textbf{464} & 211 \\
     & Blanco~\cite{blanco1997applying} & 464 & 211 \\
     & Ford~\cite{Ford2019} & 478 & 170 \\ 
     & \method(A) & 476 & 160 \\ 
     & \method(A+S) & 496 & \textbf{142} \\ 
    \bottomrule
    \end{tabular}
    \captionof{table}{
        Comparison on model size and the FLOPS required to score a single night of EEG recordings. \method{} models are significantly smaller and comparable in size/compute to baselines. 
    }
    \label{tab:efficiency}
    \vspace{-0.3cm}
\end{table}

\noindent \textbf{Analyzing Dynamic Metrics: Inference Time \& Energy} In Figure~\ref{fig:efficiency}, we benchmark the inference time and energy consumption of a Sors and \method{} model deployed on a Pixel 2 smartphone using Tensorflow Lite. We identify the following insights:
\begin{enumerate}[topsep=2mm, itemsep=0mm, parsep=1mm, leftmargin=*]
    \item \textbf{\method{} Models Run Faster} When deployed, \method{} runs $9 \times$ and $6 \times$ faster than the uncompressed model on the two datasets.
    
    \item \textbf{\method{} Models are Energy Efficient} \method{} models also consume $17\times$ and $9\times$ less energy than an uncompressed model on the Sleep-EDF and SHHS datasets, respectively.
    
    \item \textbf{Enabling Sleep Staging for Edge Computing} The above benefits demonstrate that model compression effectively translates into faster inference and a reduction in energy consumption. These benefits are crucial for deploying on the edge.
\end{enumerate}

\section{Conclusion}
We identified two key challenges in developing deep neural networks for sleep monitoring in the home environment---\textit{robustness to noise} and \textit{efficiency}.
We proposed to solve these challenges through \method{}---%
a new method that simultaneously tackles both issues.
For the sleep staging task over electroencephalogram (EEG), 
\method{} trains models that achieve up to $19 \times$ parameter reduction
and $15 \times$ MFLOPS reduction with an increase of up to 0.36 in macro-F-1 score in the presence of noise. 
By deploying these models to a smartphone,
we demonstrate that \method{} achieves up to $17 \times$ energy reduction and $9 \times$ faster inference.

\section{Acknowledgments}
This work was in part supported by the NSF award IIS-1418511, CCF-1533768, IIS-1838042, CNS-1704701, IIS-1563816; GRFP (DGE-1650044); and the National Institute of Health award NIH R01 1R01NS107291-01 and R56HL138415.

\bibliographystyle{ACM-Reference-Format}
\bibliography{main.bib}
  
\end{document}